\documentclass[twocolumn,showpacs,preprintnumbers,amsmath,amssymb]{JINST}
\usepackage{graphicx}
\usepackage{dcolumn}
\usepackage{bm}
\usepackage{epsfig}
\usepackage{amssymb}

\newcommand{\beq}[1]{\begin{equation}{\label{#1}}}
\newcommand{\eeq}[0]{\end{equation}}
\newcommand{\Fig}[1]{Figure~\ref{#1}}
\newcommand{\NIMA}{Nucl. Instr. \& Meth. {\bf A}}

\title{A multivariate approach to heavy flavour tagging with cascade training}

\author{Jo\~ao Bastos\\
  LIP-Coimbra, University of Coimbra, 3004-516 Coimbra, Portugal\\
E-mail: \email{bastos@lipc.fis.uc.pt}}

\author{Yong Liu
\thanks{Corresponding address: MS 309, 
FermiLab P. O. Box 500, Batavia, IL 60510}\\
Department of Physics and Astronomy, University of Alabama, AL 35487 \\
E-mail: \email{yongliu@fnal.gov }}

\date{\today}

\abstract{
This paper compares the performance of artificial neural networks and boosted decision trees,
with and without cascade training, for tagging b-jets in a collider experiment.
It is shown, using a Monte Carlo simulation of $WH \to l\nu q\bar{q}$ events, 
that for a b-tagging efficiency of 50\%, the light jet rejection power 
given by boosted decision trees without cascade training is about 55\% higher than that given by
artificial neural networks. The cascade training technique can improve 
the performance of boosted decision trees and artificial
neural networks at this b-tagging efficiency level 
by about 35\% and 80\% respectively. We conclude that the cascade 
trained boosted decision trees method is the
most promising technique for tagging heavy flavours at collider
experiments.    
}


\begin{document}


\section{Introduction}
\label{introduction}

Precision measurements in the top quark sector, and searches for the Higgs boson and physics beyond
the Standard Model, critically depend on the good identification (``tagging'') of jets produced by b quarks.
Tagging techniques exploit specific properties of B-hadrons to differentiate them from the large background
of jets produced by light quarks and gluons.
Due to the long lifetime of B-hadrons, the tracks from their decay form displaced vertices.
Physical observables associated to these vertices constitute the input for secondary vertex tagging.
Also, tracks from B- and D-hadron decays typically have large impact parameters, which are frequently used
to construct discriminating variables.
In a different approach, soft-lepton tagging selects leptons with low transverse momentum with respect to
the jets axis, consistent with semileptonic decays of B- and D-hadrons.
The tagging performance is substantially improved when individual taggers are combined to give a single
jet classifier.
In high energy physics, Fisher discriminants \cite{fisher} and artificial neural networks (ANN) \cite{ann-2} are
the most popular methods for combining several discriminating variables into one classifier, and have been extensively
applied to b-tagging \cite{delphi}.
Boosted decision trees (BDT) is a newly developed learning technique \cite{boo-1} that was recently introduced to high energy
experimentalists by the MiniBooNE Particle ID group \cite{boo-2}. Subsequently, it has been applied to the search for
radiative leptonic decays $B\rightarrow \gamma l \nu$ at {\sc BaBar} \cite{babar}, for detecting the first evidence of
single top quark production by the D0 collaboration \cite{d0}, and it also was proposed for supersymmetry searches
at the LHC \cite{conrad}.

The key ingredients for particle identification (PID) are the input variables, the algorithm and its internal parameters,
e.g., the number of layers, the number of hidden nodes, etc., in the case of ANN, or the number of leaves,
the number of trees, etc., in the case of BDT.
Input variables that show better separation of the distributions of signal and background events and, simultaneously,
are weakly correlated, are expected to yield greater discrimination power.
On the other hand, for a set of training events\footnote{In a pattern classification problem,
the classifier is optimized on a training data set. The predictive power of the classifier is estimated using a test data set.
These data sets must be independent of each other.} containing non-Gaussian distributed variables,
the performance of ANN is expected to be better than that of Fisher discriminants \cite{fisher}.
In turn, boosted decision trees may show superior performance than ANN \cite{boo-2}.

While the construction of improved discriminating variables and the development of more powerful algorithms are crucial activities,
we have noted that a judicious manipulation of the training sample - the cascade training technique (CTT) \cite{cascade} -
can improve the PID performance significantly.
The idea of CTT is to throw away easily identified background events and, at the same time, some signal events which
are difficult to identify. Then, one forces the algorithm to learn the feature differences between signal and
background from the remaining signal and background events. To some extent, the premise behind CTT is that,
once the algorithm learns to separate background events which
are difficult to identify from signal events, it will be able to separate those easily identified background
events from signal events naturally.
In other words, the CTT can help the algorithm exhaust the difference of features between signal and background events. 
The complete CTT procedure has been summarized in \cite{cascade}. Concretely, one needs to split the signal/background
event sample into three sets, say, $A^S/A^B$, $B^S/B^B$ and  $C^S/C^B$. Here, the superscript $S/B$ denotes signal/background.
Then, one trains the learning algorithm with the training samples $A^S$ and $A^B$ and looks at the performance with 
the test samples $B^S$ and $B^B$, based on which one determines the cut value to select
training events from samples $C^S/C^B$ for the second training step. After training the algorithm on the selected events
of samples $C^S/C^B$, one tests it again on samples $B^S$ and $B^B$. 
The first training step only serves for selection of the training events for the second training step,
while the algorithm built in the second training step is the one that is actually used for PID purposes.

It has been shown, in the context of the MiniBooNE experiment, that the CTT procedure can improve significantly
the PID performance in the very low background contamination region.
As a general training procedure, the CTT may be applied to any PID task which involves multivariate analysis techniques.

The goals of this paper are two-fold. First, we show that BDT performs
better than ANN for tagging heavy flavours in collider experiments.
Second, we show that the CTT procedure has a large impact on the
performances of both BDT and ANN.
For this purpose, we generated a sample of $WH$ events and reconstructed several discriminating variables typically
employed in b-tagging techniques.
The next section gives a brief description of the Monte Carlo simulation used in this study.
Section \ref{discriminant_variables} describes the discriminating variables which feed the tagging algorithms.
The performances of boosted decision trees, artificial neural networks, with and without the cascade training technique,
are compared in Section \ref{results}.
Finally, conclusions are given in Section~\ref{conclusions}.


\section{Monte Carlo simulation}
\label{monte_carlo_simulation}

The studies described in this paper were done with events generated with PYTHIA 6.319 \cite{pythia}.
We considered the environment of the LHC collider, in which $pp$ interactions with a center-of-mass energy of
14 TeV are produced.
One of the benchmark channels for b-tagging studies at the LHC is the associated $WH$ production.
We generated $WH$ events with $m_H$ = 120 GeV/c$^2$, the $W$ boson decaying semileptonically
$W\to l\nu$ and the Higgs boson decaying to quark pairs $H\to q\bar{q}$.
Initial and final state radiation and multiple interactions were included in the simulation.
No minimum bias interactions superimposed on the hard-scattering were included in the simulation.
The effect of pile-up interactions in the performance of the algorithms is discussed in Section~4.1.

Tracks are parametrized by the following set of 5 parameters: $d_0$, $z_0$, $\phi$, $\cot\theta$ and $1/p_T$.
The transverse impact parameter $d_0$ is the distance of closest approach of the track to the primary vertex
in the plane perpendicular to the beam-line. The longitudinal impact parameter $z_0$ is the component along
the beam-line of the distance of closest approach. The parameters $\phi$ and $\theta$ are the azimuthal
and polar angles of the track, respectively, and $1/p_T$ is the inverse of the particle transverse momentum.
In order to simulate measurement errors, these parameters were smeared with Gaussian resolution functions.
The transverse and longitudinal impact parameters were smeared with standard deviations
$\sigma_{d_0} = 10$~$\mu$m and $\sigma_{z_0} = 100$~$\mu$m,
the angle $\phi$ with $\sigma_{\phi} = 0.10$~mrad, $\cot\theta$ with $\sigma_{\cot\theta} = 0.001$ and the
inverse of the transverse momentum with $\sigma_{1/p_T}= 0.001$ GeV$^{-1}$. The primary vertex positions were smeared
with Gaussian resolution functions with $\sigma_x = \sigma_y = 50$~$\mu$m and $\sigma_z = 100$~$\mu$m.
A jet is formed by all stable charged particles inside a cone $\Delta R  = \sqrt{(\Delta\phi)^2 + (\Delta\eta)^2} < 0.4$
around its axis, where $\eta = -\log\left(\tan(\theta / 2)\right)$ is the track pseudorapidity.


\section{Discriminating variables}
\label{discriminant_variables}

The physical observables used for discrimination between b-jets and light jets are taken from well known
``spatial" b-tagging techniques.
In order to simulate the typical acceptance of collider experiments, only jets with $p_T > 15$ GeV/c and
$|\eta| < 2.5$ were considered as taggable.

\subsection{Impact parameter tag}
\label{impact_parameter_tag}

\begin{figure}[htb]
\centering
\epsfig{file=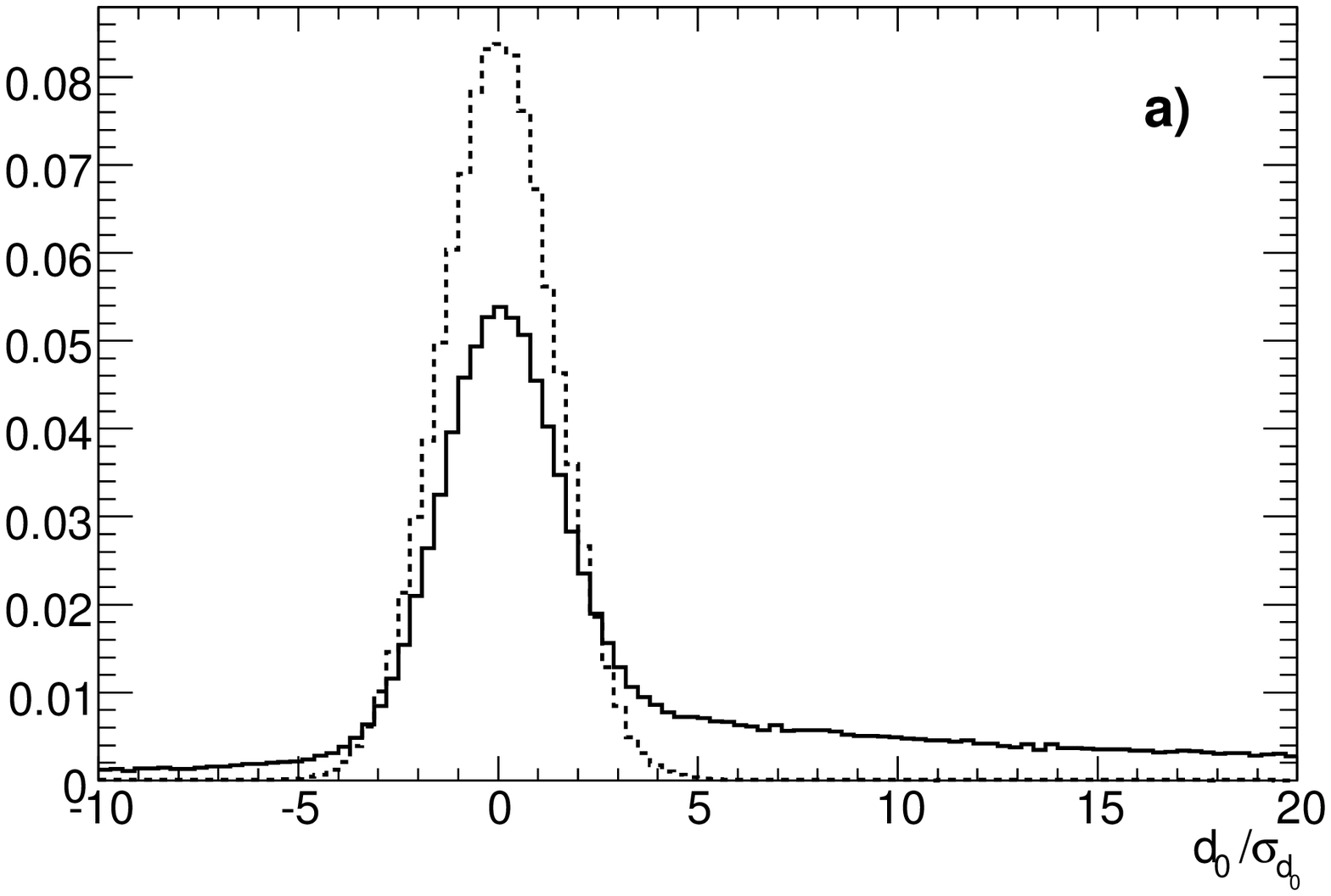, width=8cm} \\
\epsfig{file=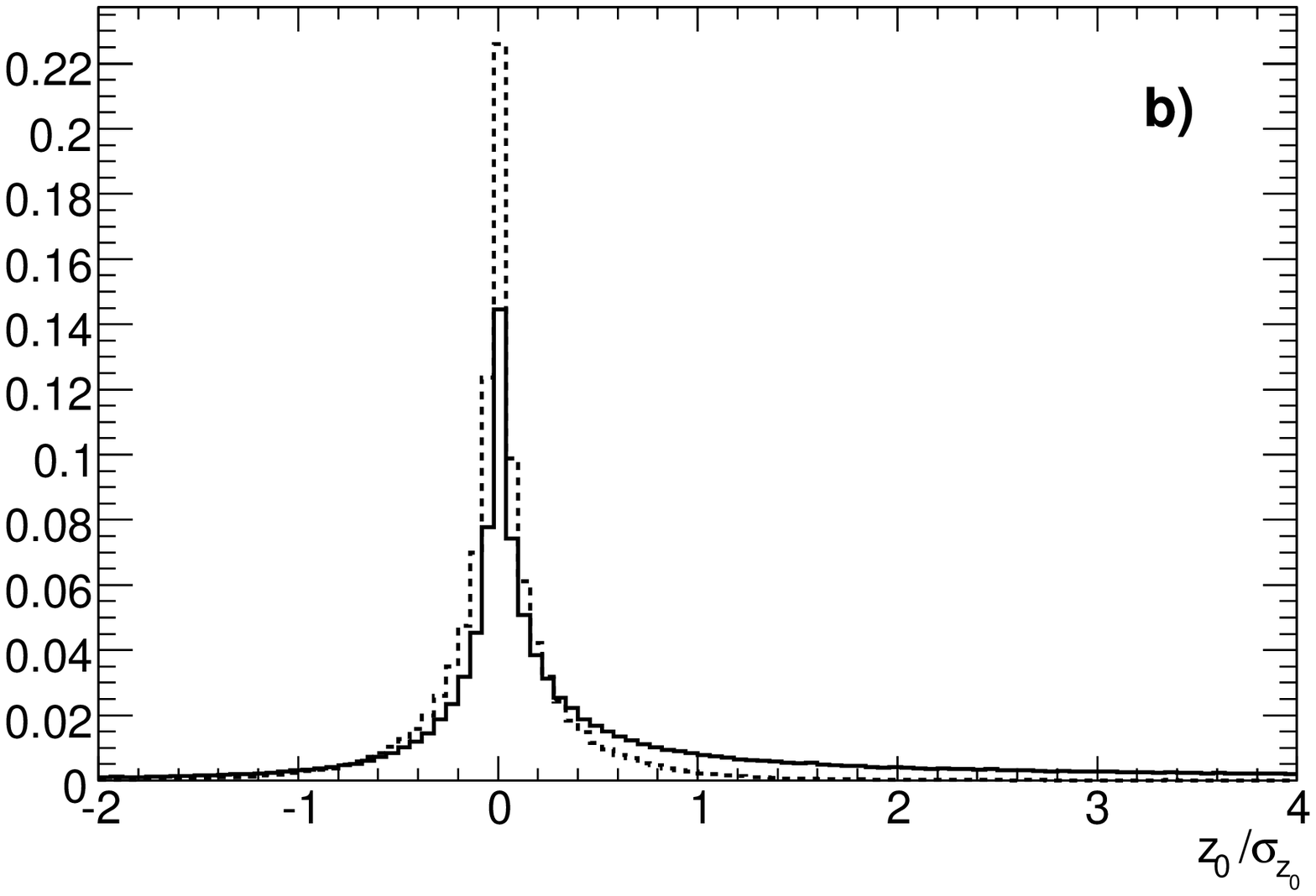, width=8cm}
\caption{Normalized distributions of (a) transverse and (b) longitudinal impact parameter significances for tracks
in b-jets (solid line) and u-jets (dashed line).}
\label{ipsig}
\end{figure}

Due to the long decay distances traveled by B-hadrons, tracks from b-jets have on average larger impact parameters
than tracks from light jets, since sizeable impact parameters in light jets are exclusively due to measurement errors,
$V^0$ decays, conversions and hadronic interactions in the detector material.
Therefore, the impact parameter of jet tracks can be used to build a useful variable for discrimination between
b-jets and light jets.
\Fig{ipsig} shows the normalized distributions of (a) signed transverse impact parameter significances $S_{d_0} = d_0 / \sigma_{d_0}$
and (b) signed longitudinal impact parameter significances $S_{z_0} = z_0 / \sigma_{z_0}$ of tracks in b-jets (solid line)
and u-jets (dashed line).
A positive (negative) sign is assigned to the impact parameter if the track intersects the jet axis in front (behind)
of the primary vertex.
These distributions give likelihood functions $b(S)$ and $u(S)$ for a track to belong to a
b-jet and a u-jet, respectively.
A jet weight is defined as the sum of the log-likelihood ratio over all tracks in the jet:
\beq{weight}
   w_{jet} = \sum_{i\in jet} \ln \left(\frac{b(S_i)}{u(S_i)}\right)\,.
\eeq

\begin{figure}[htb]
\centering
\epsfig{file=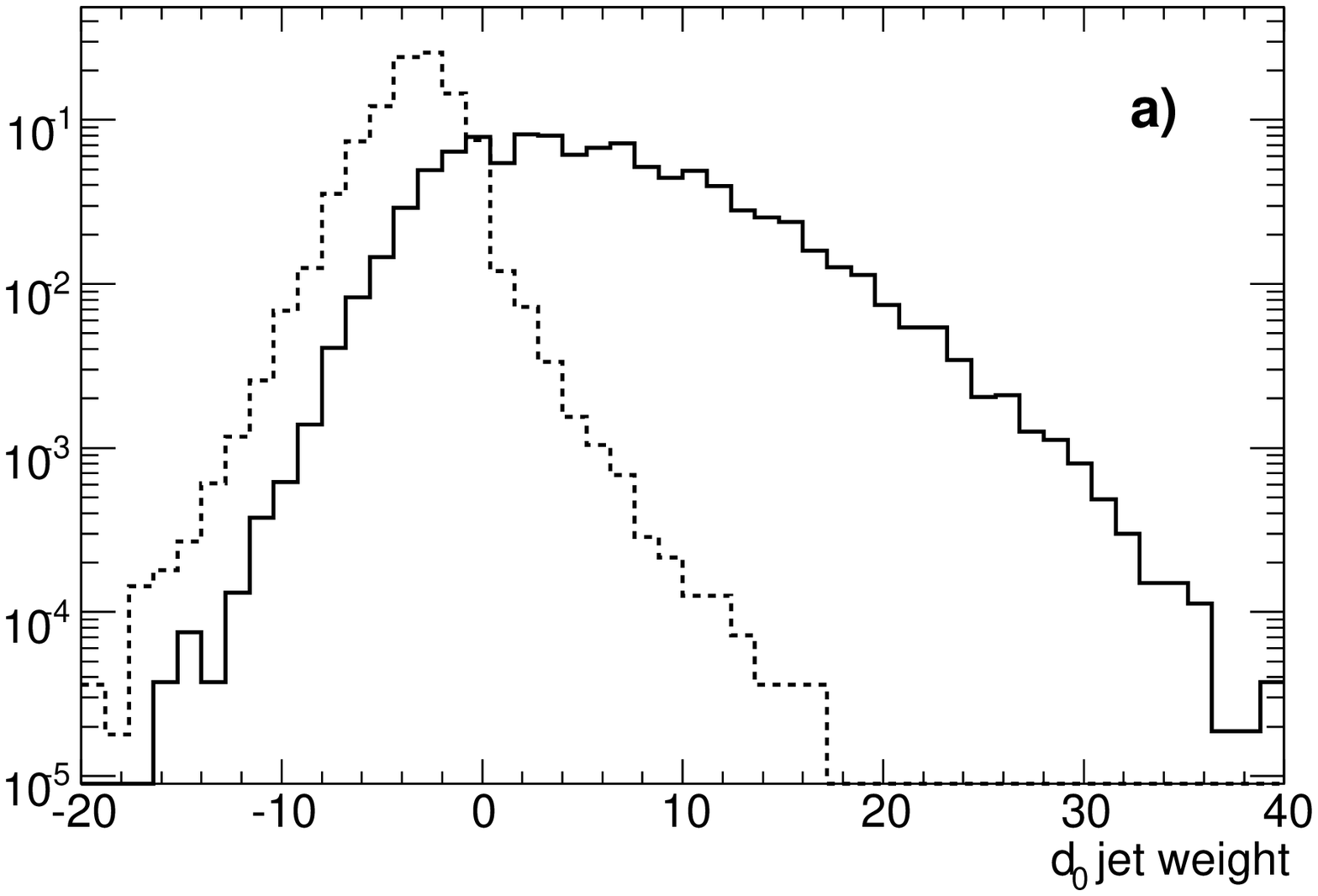, width=8cm} \\
\epsfig{file=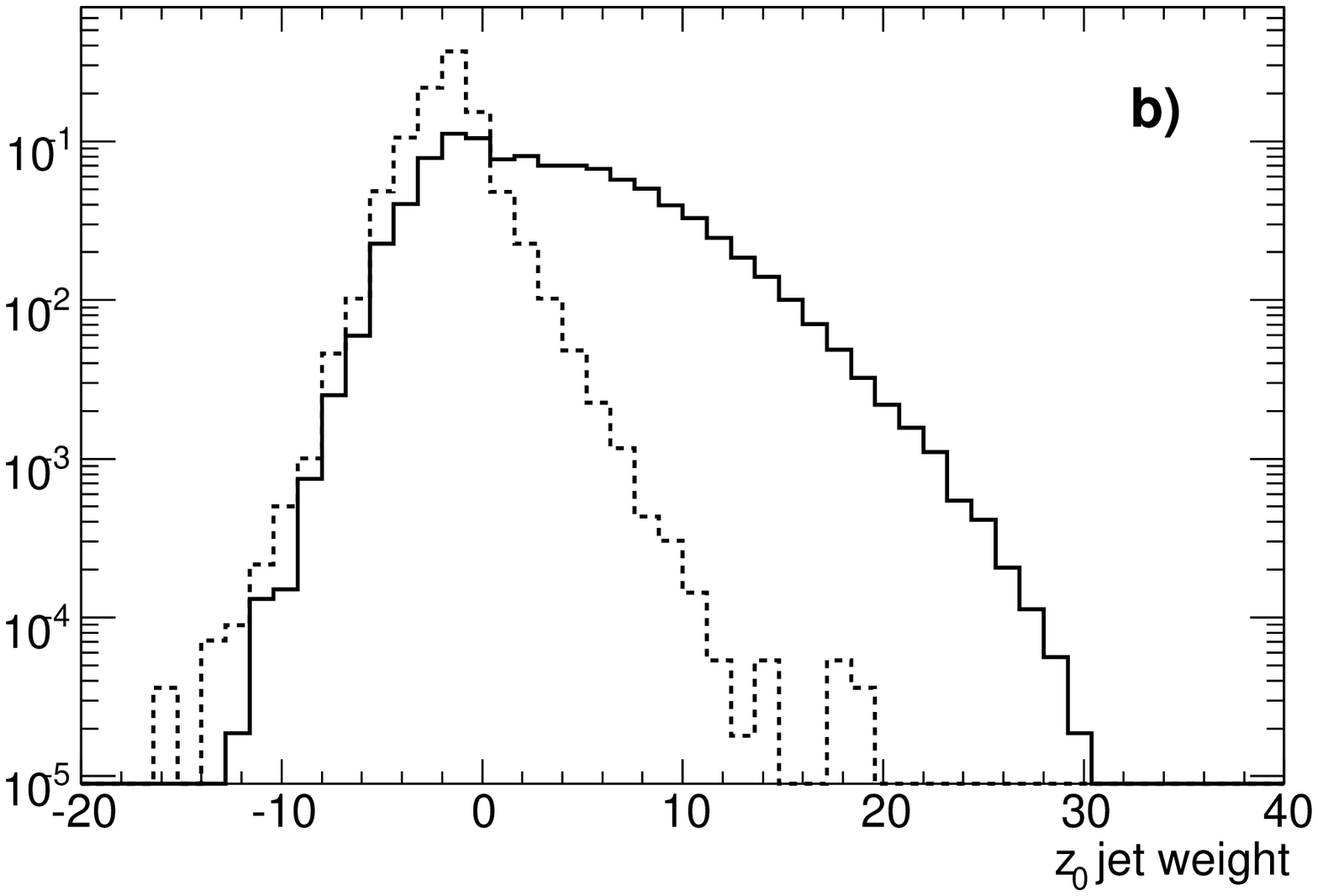, width=8cm}
\caption{Jet weight normalized distributions given by the transverse impact parameter (a) and longitudinal impact
parameter (b). The solid (dashed) line corresponds to b-jets (u-jets).}
\label{weights}
\end{figure}

\Fig{weights} shows the normalized distribution of jet weights for u and b quarks.
Because the transverse impact parameter has better resolution, it yields greater discrimination power.
A given efficiency for selecting b-jets is obtained by selecting jets with weights above some
threshold level. Obviously, for moderate or high selection efficiencies there will always be some contamination
with light jets.

\subsection{Secondary vertex tag}
\label{secondary_vertex_tag}

\begin{figure}
\centering
\epsfig{file=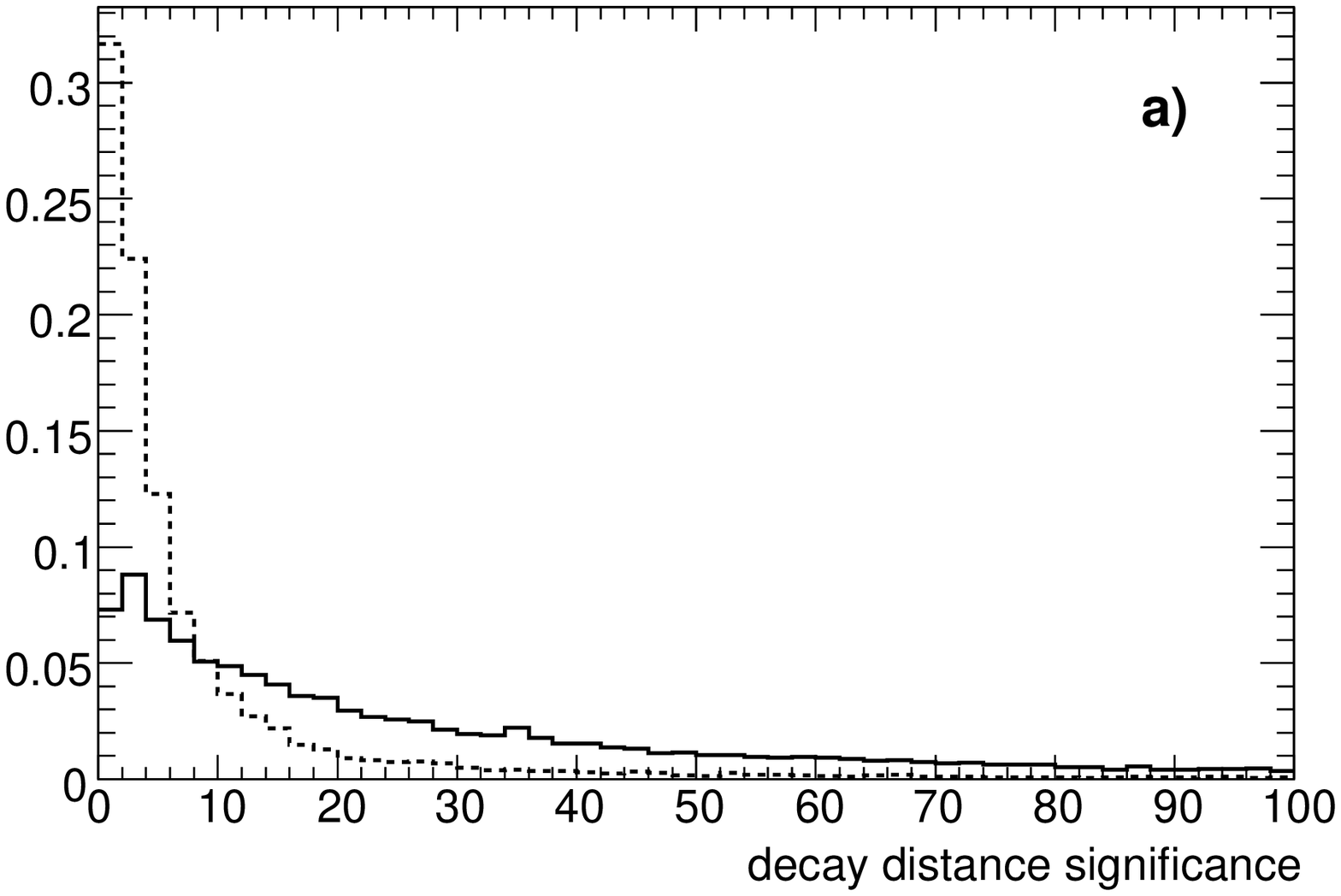,      width=8cm} \\
\epsfig{file=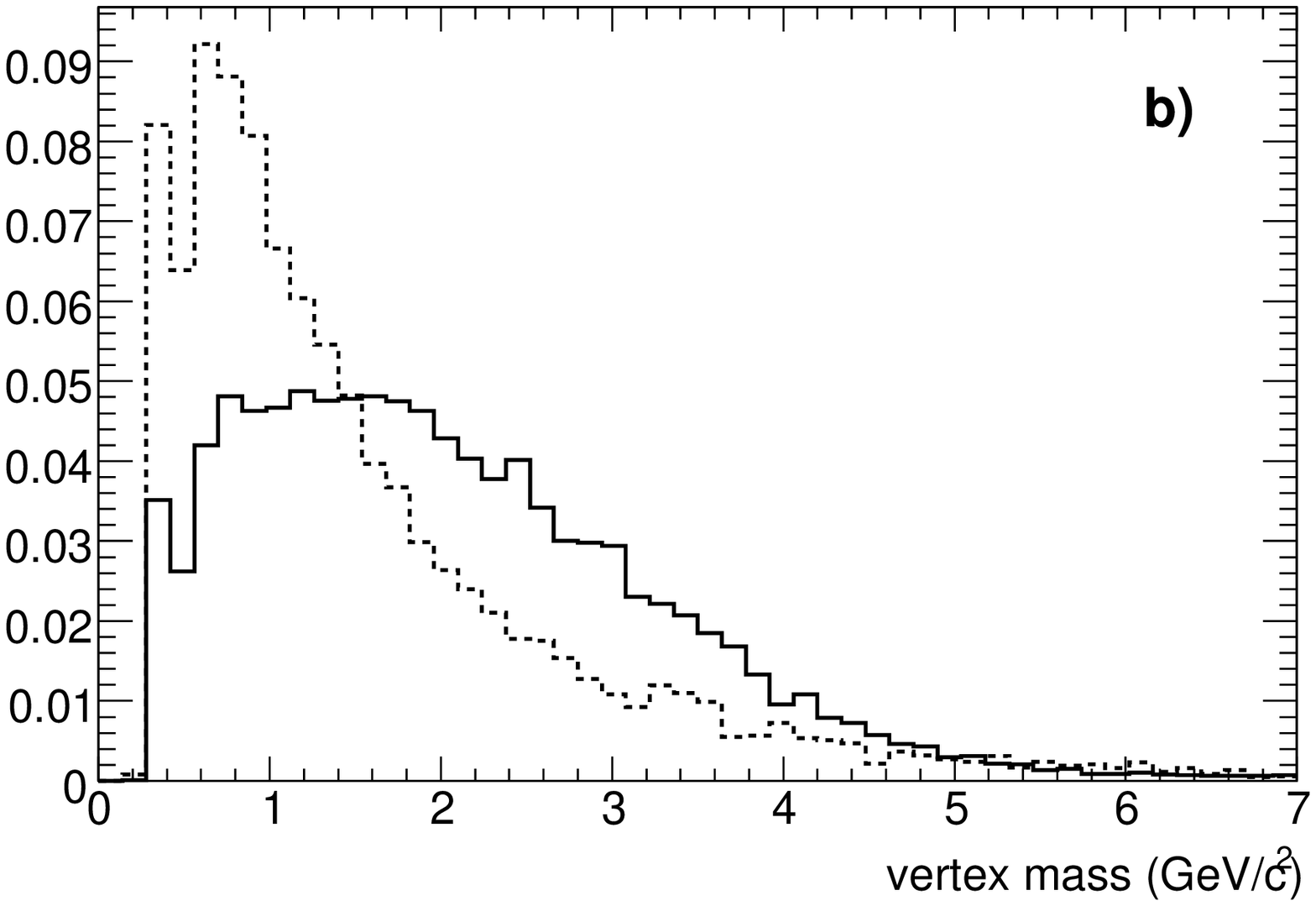, width=8cm} \\
\epsfig{file=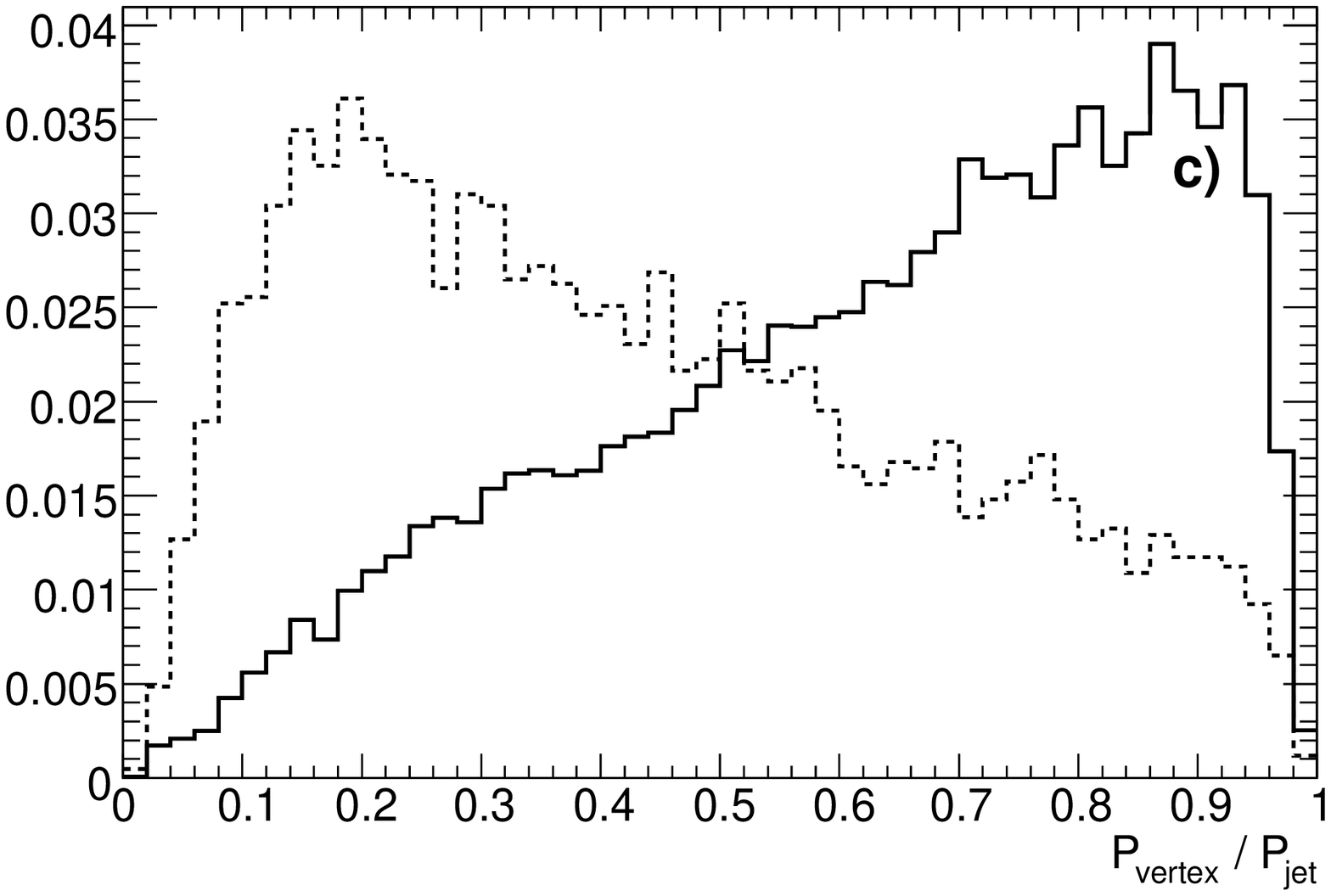,   width=8cm}
\caption{Normalized distributions of (a) decay distance significance of the secondary vertex, (b) invariant mass of
tracks associated to the secondary vertex, (c) fraction of jet momentum in the secondary vertex. The solid (dashed)
line corresponds to b-jets (u-jets).}
\label{vertex}
\end{figure}

An alternative approach for building b-tagging discriminating variables reconstructs displaced
secondary vertices from B- and D-hadron decays inside the jet. Secondary vertices were reconstructed with
Billoir and Qian's fast vertex fitting algorithm \cite{billoir_qian}.
For purposes of secondary vertex b-tagging the exact topology of the secondary vertex is irrelevant and,
therefore, an inclusive vertex search is performed.
All jet tracks with large transverse impact parameter significance participate in the vertex fit
and vertices compatible with $V^0$ decays are rejected.
\Fig{vertex}(a) shows the decay distance significance for b-jets and u-jets.
Besides the decay distance significance, other variables associated to the secondary vertex may have discrimination
power, such as the vertex mass (\Fig{vertex}(b)) and the ratio between the absolute momentum sum of tracks in the
secondary vertex and that of all tracks in the jet (\Fig{vertex}(c)).

\subsection{One-prong tag}
\label{one_prong_tag}

\begin{figure}[htb]
\centering
\epsfig{file=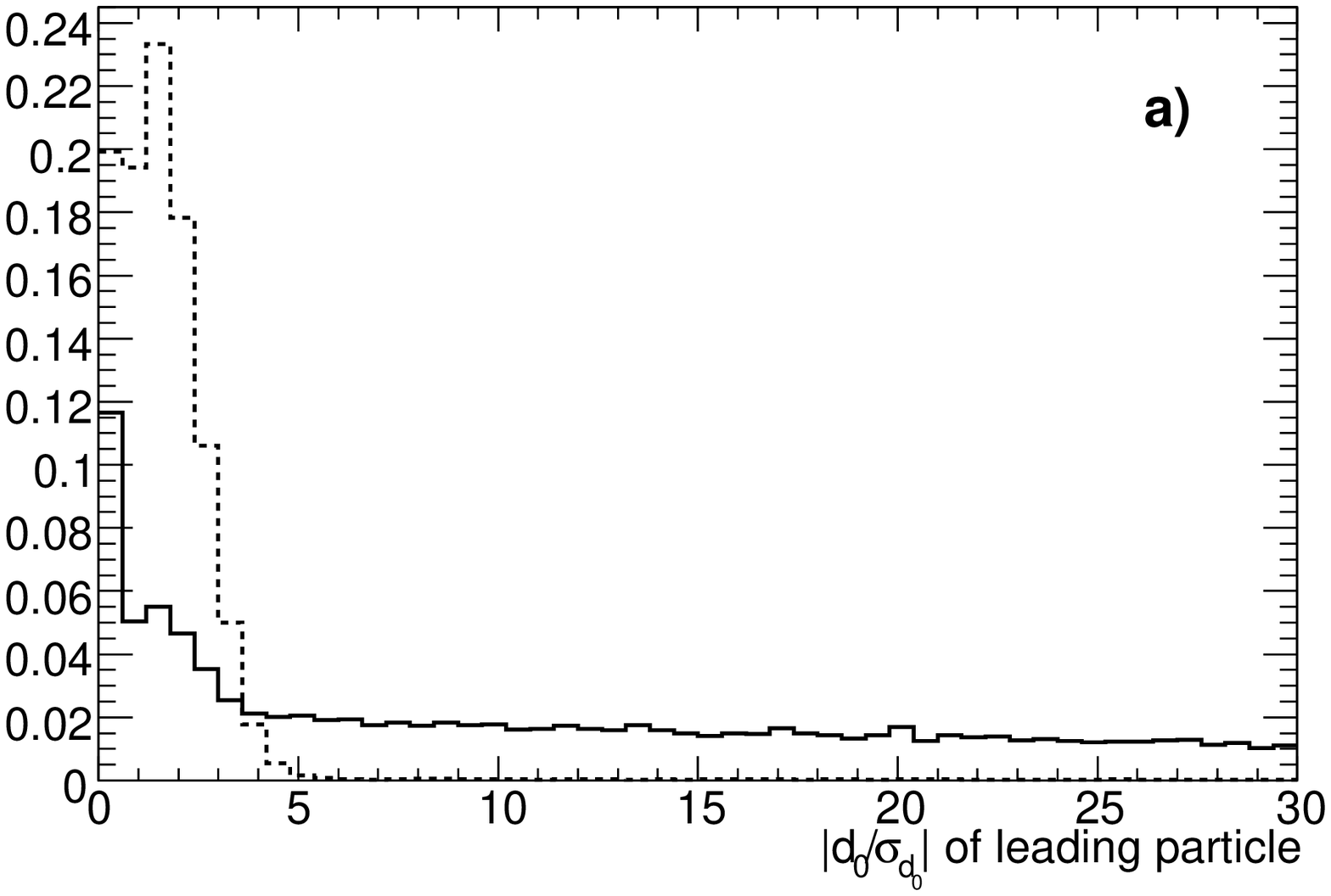, width=8cm} \\
\epsfig{file=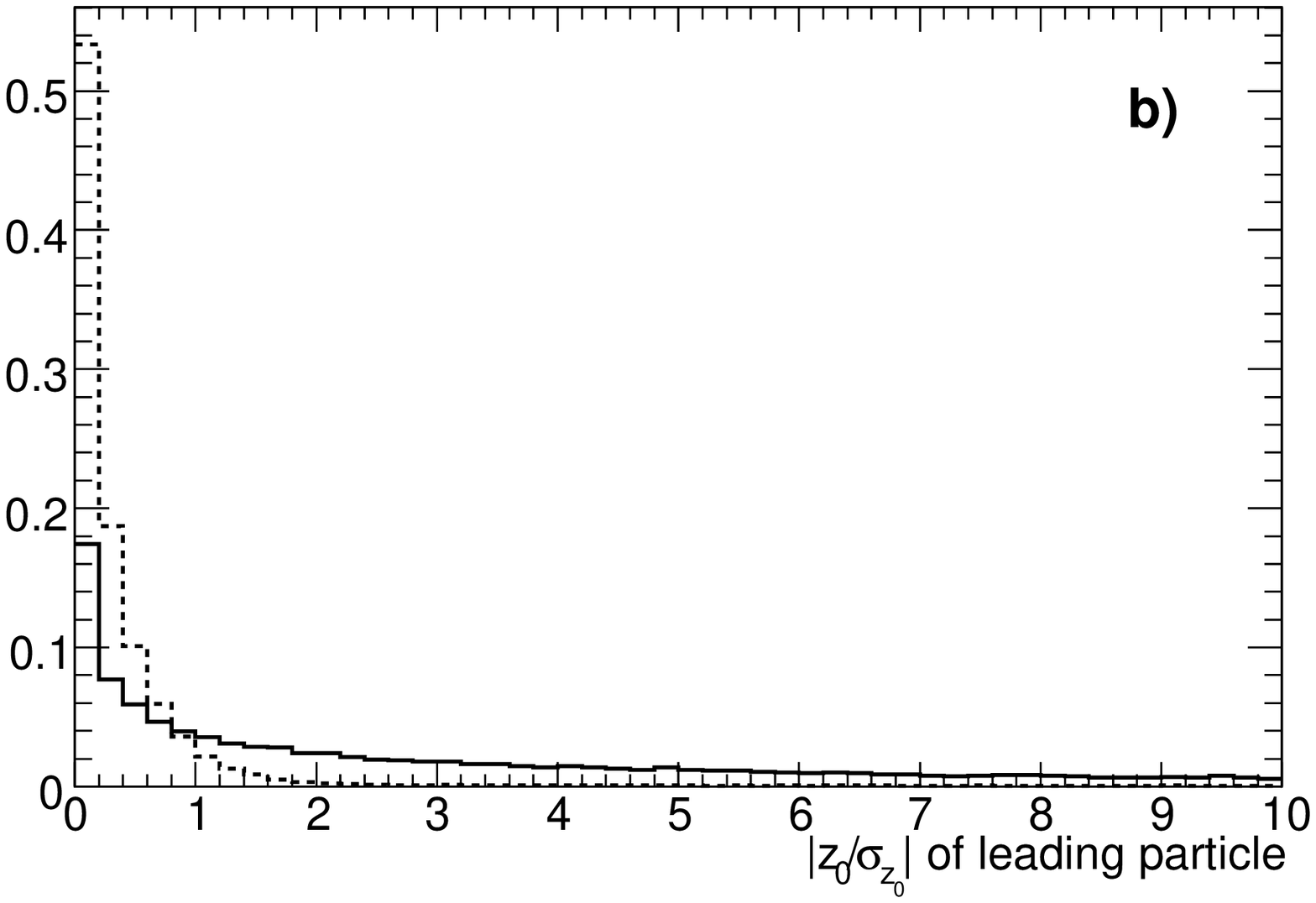, width=8cm}
\caption{Normalized distributions of the maximal (a) transverse and (b) longitudinal impact parameter significances
in jets. The solid (dashed) line corresponds to b-jets (u-jets).}
\label{maxip}
\end{figure}

For one-prong decays of B- and D-hadrons the secondary vertex fit fails.
In this situation, though, some information can still be extracted from tracks in the jet.
For instance, the maximal transverse and longitudinal impact parameters of jet tracks clearly have
discrimination power, as can be observed in \Fig{maxip}.

\section{Results}
\label{results}

In order to perform the studies described below, a sample of 200,000 b-jets and 200,000 u-jets from Higgs boson decays
were selected from the generated events.
Considering jets from the same physics channel, allows us to infer the performance of the algorithms for separating b-jets
from background jets with similar kinematics.
Each jet was characterized by 7 attributes corresponding to the 7 discriminating variables described in the previous
section (jet weight distributions for transverse and longitudinal impact parameters, decay distance significance
of the secondary vertex, invariant mass of tracks associated to the secondary vertex, fraction of jet momentum in the
secondary vertex, and the maximal transverse and longitudinal impact parameter significances of tracks in the jet).

The BDT was implemented using the StatPatternRecognition package \cite{narsk}. 
The reader is referred to the StatPatternRecognition manual for a description of the BDT algorithm employed here.
The best performance were obtained with about 500 trees and a minimum number of jets per leaf of about 7000.
In order to implement the ANN, the Jetnet 3.0 package \cite{ann-1} was considered, since it has been broadly accepted and
used in leading high energy physics experiments since the 1990's \cite{ann-2,delphi}.
The architecture of the network consisted of 7 nodes in the input layer (corresponding to the 7 discriminating variables
mentioned above), 14 (15) nodes in the hidden layer for the first (second) training step and 1 node in the output layer.
The network was trained with a learning rate parameter $\eta = 0.8$ and a momentum parameter $\alpha=0.5$.
The number of epochs (training cycles) was 100.
For both ANN and BDT, the first 25,000 b-jets and 25,000 u-jets in the data were used in
the first training step. Then, the following 40,000 b-jets and 40,000 u-jets were used for
testing the algorithms.
From the remaining jets, those with an output score greater
than 0.4 for ANN and 0.15 for BDT were used in the second training step. These threshold scores 
represent a compromise between the better performance obtained when higher values are considered 
and having enough u-jets for the second training step.

\begin{figure}[htb]
\centering
\includegraphics[width=8.0cm,height=9.0cm]{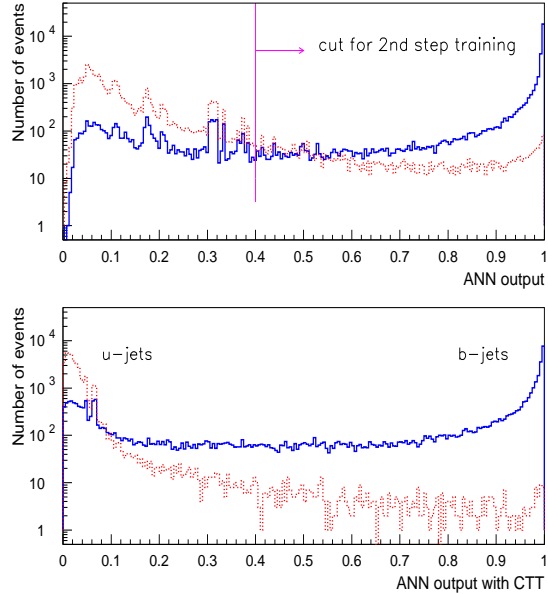}
\caption{Output distribution for jets in the test sample, given by the artificial neural network. 
The top (bottom) plot corresponds to the first (second) training step.
The solid (dashed) line corresponds to b-jets (u-jets).
The vertical line at 0.4 shows the cut used to select training jets for the second training step.
}
\label{anndis}
\end{figure}

\begin{figure}[htb]
\centering
\includegraphics[width=8.0cm,height=9.0cm]{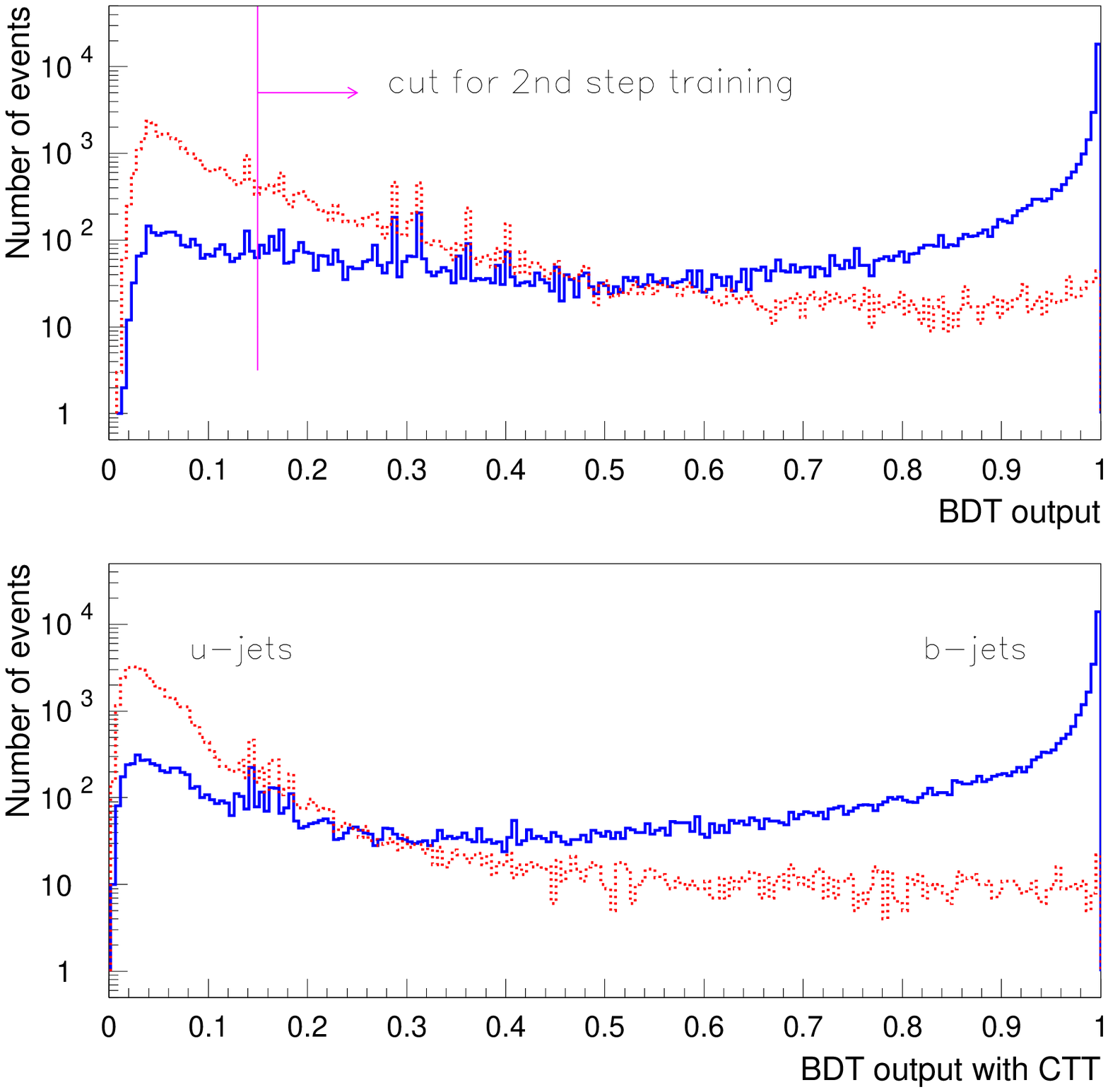}
\caption{Output distribution for jets in the test sample, given by boosted decision trees.
The top (bottom) plot corresponds to the first (second) training step.
The solid (dashed) line corresponds to b-jets (u-jets).
The vertical line at 0.15 shows the cut used to select training jets for the second training step.
}
\label{bdtdis}
\end{figure}

The ANN and BDT output distributions for jets in the test sample are shown in \Fig{anndis} and \Fig{bdtdis},
respectively. The top panels show the output after the first training step, while the bottom panels correspond to the
output obtained after the second training step.
Also shown in the top panels of \Fig{anndis} and \Fig{bdtdis} is a vertical line that represents the cut 
to select training jets for the second training step.
The solid lines correspond to b-jets in the test sample, while the dashed lines correspond to u-jets.
From \Fig{anndis} and \Fig{bdtdis}, one can see that, the ratio between the number of b-jets and the
number of u-jets gets larger in those very right end (signal region) bins, but smaller in very left end (background region) bins,
after the second training step.

\begin{figure}[htb]
\centering
\includegraphics[width=9.0cm,height=9.0cm]{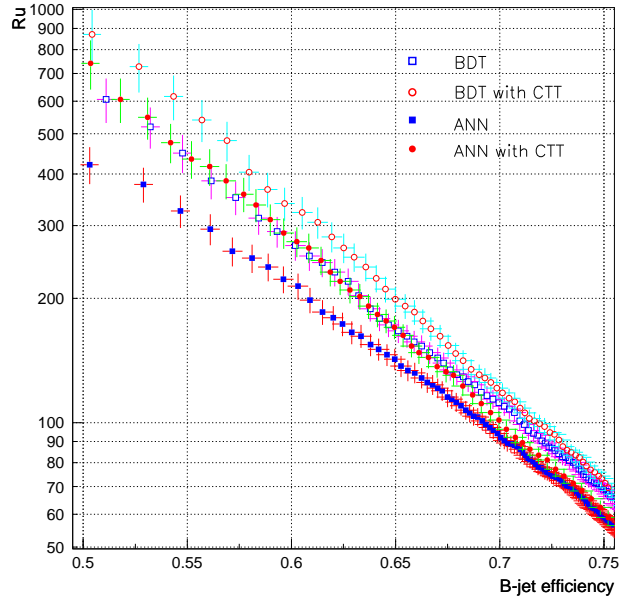}
\caption{Light jet rejection power versus b-tagging efficiency.
Open squares:  first step BDT.
Open circles:  second step BDT.
Solid squares:  first step ANN.
Solid points:  second step ANN.
}
\label{rejeff}
\end{figure}

Jets with an output above some specified threshold value are tagged as b-jets, while jets with a score below this value are
tagged as light jets. The threshold value is contingent on the desired efficiency for tagging b-jets
$\varepsilon_b = N_b^{tag} / N_b$, where $N_b$ is the number of b-jets in the data and $N_b^{tag}$ is the number of
tagged b-jets, or, alternatively, on the tolerated level of contamination by light jets.
The light jet rejection power $R_u = \varepsilon_u^{-1}$ versus b-tagging efficiency $\varepsilon_b$ is shown
in \Fig{rejeff}. We can see that, above 70\% b-tagging efficiency, there are no significant differences in the performances
of ANN and BDT, irrespective of using CTT. However, the light jet rejection power is very poor.
On the other hand, for lower b-tagging efficiencies, significant differences in the performances are observed and,
the smaller the b-tagging efficiency is, the larger are the differences between ANN and BDT, with and without CTT.

For any level of b-tagging efficiency, BDT outperforms ANN. However, with CTT, ANN shows a comparable
performance to that of BDT without CTT.
The light jet rejection power for four b-tagging efficiency levels is shown in Table 1.
For a b-tagging efficiency of 60\%, the light jet rejection power given by BDT is about 25\% better than that
given by ANN, while CTT gives an improvement of about 30\% for ANN and an improvement of about 20\% for BDT.
For a b-tagging efficiency of 50\%, the light jet rejection power given by BDT is about 55\% better than that given by ANN,
and CTT can improve the performance of ANN by more than 80\% and that of BDT by about 35\%.
Finally, for a b-tagging efficiency of 40\%, CTT almost doubles the light jet rejection power given by ANN.

\begin{table}[htb]
\begin{center}
{
\begin{tabular}{|l|c|c|c|c|}
\hline
b-tagging efficiency &   40\% & 50\%  &  60\% &  70\%  \\ \hline\hline
BDT                  & 1333 $\pm$ 244 & 667 $\pm$  86 & 274 $\pm$ 23 & 112 $\pm$ 6 \\ \hline
BDT with CTT         & 1482 $\pm$ 285 & 889 $\pm$ 133 & 331 $\pm$ 30 & 120 $\pm$ 7 \\ \hline\hline
ANN                  &  588 $\pm$  71 & 426 $\pm$  44 & 217 $\pm$ 16 &  93 $\pm$ 5 \\ \hline
ANN with CTT         & 1143 $\pm$ 193 & 769 $\pm$ 107 & 282 $\pm$ 24 & 101 $\pm$ 5 \\ \hline 
\end{tabular}
}
\vspace{0mm}
\caption{Light jet rejection power for four b-tagging efficiency levels, for boosted decision trees
and artificial neural networks, with and without the cascade training technique.
The errors given in the tables herein and after are statistical errors.}
\end{center}
\label{efftab1}
\end{table}


\subsection{Robustness of the results}
\label{robustness}

The effect of pile-up interactions on these results was studied.
To accomplish this, minimum bias interactions were superimposed on the hard-scattering in the simulation.
The minimum bias interactions included low-$p_T$ pp interactions, single and double diffractive interactions and
elastic scatterings.
Three luminosities per bunch-bunch crossing were considered: 0.015, 0.030 and 0.045 mb$^{-1}$.
For the LHC bunch crossing rate of 40 MHz and 100 days of data acquisition, these values correspond
to integrated luminosities of about 5, 10 and 15 fb$^{-1}$ per year, respectively.
Table 2 shows the light jet rejection power for a b-tagging efficiency of 60\%.
The presence of minimum bias interactions degrades the performance of all algorithms.
Higher luminosities result in lower performance since more pile-up tracks are selected in the analysis.
For a bunch-bunch crossing luminosity of 0.045 mb$^{-1}$ the rejection power of the algorithms is decreased
by about 20\% with respect to the no pile-up scenario.
Regardless of the luminosity value, the method that shows the best performance is BDT with cascade training.

\begin{table}[htb]
\begin{center}
{
\begin{tabular}{|l|c|c|c|c|}
\hline
                     &   no pile-up & 0.015 mb$^{-1}$ &  0.030 mb$^{-1}$ & 0.045 mb$^{-1}$  \\ \hline\hline
BDT                  & 274 $\pm$ 23 &   253 $\pm$ 20  &     250 $\pm$ 19 &     219 $\pm$ 16 \\ \hline
BDT with CTT         & 331 $\pm$ 30 &   294 $\pm$ 25  &     274 $\pm$ 23 &     270 $\pm$ 22 \\ \hline\hline
ANN                  & 217 $\pm$ 16 &   191 $\pm$ 13  &     189 $\pm$ 13 &     175 $\pm$ 12 \\ \hline
ANN with CTT         & 282 $\pm$ 24 &   242 $\pm$ 19  &     240 $\pm$ 19 &     227 $\pm$ 17 \\ \hline 
\end{tabular}
}
\vspace{0mm}
\caption{Light jet rejection power for 60\% b-tagging efficiency, for boosted decision trees
and artificial neural networks, with and without the cascade training technique, when
minimum bias interactions are superimposed on the hard-scattering and for three different values
of the bunch-bunch crossing luminosity.}
\end{center}
\label{efftab2}
\end{table}


The performance of the algorithms as a function of the detector resolution was investigated by varying
the widths of the Gaussian resolution functions with which the track parameters are smeared.
Scenarios of worse (better) detector resolutions were obtained by increasing (decreasing) all widths by 25\% and 50\%.
The performance of the algorithms can be found in Table 3, for a b-tagging efficiency of 60\%.
As expected, worse resolutions degrade the performance of all methods, since the input variables lose
discriminating power.
A decrease in the performance of up to about 25\% can be observed when the widths of the resolution functions are
increased by 50\%. An improvement in the performance of up to about 20\% is obtained when the widths of the resolution
functions are decreased by 50\%. Again, BDT with cascade training shows the best performance in all scenarios.

\begin{table}[htb]
\begin{center}
{
\begin{tabular}{|l|c|c|c|c|c|}
\hline
                     &         -50\% &        -25\% &          0\% &        +25\% &         +50\%  \\ \hline\hline
BDT                  & 333 $\pm$ 31  & 286 $\pm$ 24 & 274 $\pm$ 23 & 245 $\pm$ 19 & 220  $\pm$ 16  \\ \hline
BDT with CTT         & 370 $\pm$ 36  & 351 $\pm$ 33 & 331 $\pm$ 30 & 265 $\pm$ 22 & 242  $\pm$ 19  \\ \hline\hline
ANN                  & 255 $\pm$ 20  & 216 $\pm$ 16 & 217 $\pm$ 16 & 192 $\pm$ 13 & 179  $\pm$ 12  \\ \hline
ANN with CTT         & 305 $\pm$ 27  & 276 $\pm$ 23 & 282 $\pm$ 24 & 231 $\pm$ 18 & 205  $\pm$ 15  \\ \hline 
\end{tabular}
}
\vspace{0mm}
\caption{Light jet rejection power for 60\% b-tagging efficiency, for boosted decision trees
and artificial neural networks, with and without the cascade training technique, for different
detector resolutions.}
\end{center}
\label{efftab3}
\end{table}


So far it was assumed that all tracks are reconstructed with an efficiency of 100\%.
Since in a typical collider experiment the track reconstruction efficiency is smaller than 100\%,
it is important to investigate how reduced track reconstruction efficiencies affect the performance of the algorithms.
In order to simulate reconstruction inefficiencies, given a track reconstruction efficiency of $\varepsilon$,
tracks were randomly removed from the events with probability
\mbox{$1 - \varepsilon$}.\footnote{Two simplifications are made here: the reconstruction efficiency is
independent of the particle type and it is constant with respect to $p_T$ and $\eta$.}
Reduced track reconstruction efficiencies result in a lower number of tracks available for calculating
the jet weights (Equation 3.1) and in a lower number of tracks participating in the secondary vertex fit.
The rejection power of the algorithms for track reconstruction efficiencies of 95\% and 90\%, and
for a b-tagging efficiency of 60\%, are shown in Table 4. Lower
track reconstruction efficiencies degrade substantially the performance of all algorithms.
For a track reconstruction efficiency of 90\%, the rejection power can be up to about 20\% smaller when compared
with the scenario of a track reconstruction efficiency of 100\%.

\begin{table}[htb]
\begin{center}
{
\begin{tabular}{|l|c|c|c|}
\hline
                     &        100\% &         95\% &        90\%  \\ \hline\hline
BDT                  & 274 $\pm$ 33 & 261 $\pm$ 21 & 221 $\pm$ 17  \\ \hline
BDT with CTT         & 331 $\pm$ 30 & 325 $\pm$ 29 & 278 $\pm$ 23  \\ \hline\hline
ANN                  & 217 $\pm$ 16 & 208 $\pm$ 15 & 186 $\pm$ 13  \\ \hline
ANN with CTT         & 281 $\pm$ 24 & 242 $\pm$ 19 & 221 $\pm$ 17  \\ \hline 
\end{tabular}
}
\vspace{0mm}
\caption{Light jet rejection power for 60\% b-tagging efficiency, for boosted decision trees
and artificial neural networks, with and without the cascade training technique, for different
track reconstruction efficiencies.}
\end{center}
\label{efftab4}
\end{table}


Finally, the dependence of the results on the size of the training data sample was investigated.
The original data sample consisted of 200k b-jets and 200k u-jets. Both samples were divided
in three subsamples of 25k jets for the first training step, 40k jets for testing and 135k jets for the
second training step.
Here, three additional sizes for the training subsamples are considered: 15k, 55k and 75k jets
for the first training step and 75k, 255k and 315k jets for the second training step.
The subsample for testing consists of 40k jets for all samples.
Table 5 shows the rejections power for different sizes of the training samples.
Excluding ANN with CTT, which seems to saturate for larger training samples, 
all other algorithms show better performance for larger samples.
The increase in performance is greater for BDT. For this algorithm the rejection power increases
by about 35\% when the training samples are increased from 15k and 75k jets to 75k and 315k jets
for the first and second step training respectively. 
The reason why the performance of some algorithms saturate, 
while the performance of others do not saturate is still under investigation.

\begin{table}[htb]
\begin{center}
{
\begin{tabular}{|l|c|c|c|c|}
\hline
                     &      15k+75k &     25k+135k &     55k+255k &     75k+315k  \\ \hline\hline
BDT                  & 253 $\pm$ 20 & 274 $\pm$ 33 & 325 $\pm$ 29 & 345 $\pm$ 32  \\ \hline
BDT with CTT         & 301 $\pm$ 26 & 331 $\pm$ 30 & 354 $\pm$ 33 & 367 $\pm$ 35  \\ \hline\hline
ANN                  & 199 $\pm$ 14 & 217 $\pm$ 16 & 233 $\pm$ 18 & 244 $\pm$ 19  \\ \hline
ANN with CTT         & 255 $\pm$ 20 & 281 $\pm$ 24 & 288 $\pm$ 24 & 282 $\pm$ 22  \\ \hline 
\end{tabular}
}
\vspace{0mm}
\caption{Light jet rejection power for 60\% b-tagging efficiency, for boosted decision trees
and artificial neural networks, with and without the cascade training technique, for different
sizes of the training samples.}
\end{center}
\label{efftab5}
\end{table}


\section{Conclusions}
\label{conclusions}

The studies presented in this paper indicate that boosted decision trees outperform neural networks for
tagging b-jets, using a Monte Carlo simulation of $WH \to l\nu q\bar{q}$ events, and sensible physical
observables as discriminating variables.
For a b-tagging efficiency of 60\%, the light jet rejection given by boosted decision trees is about 25\% higher
than that given by artificial neural networks. For lower b-tagging efficiencies this difference is even larger.
Furthermore, the cascade training technique improves the performance of both methods: for a b-tagging efficiency of
60\%, a 30\% improvement is observed for artificial neural networks, while a 20\% improvement is observed for
boosted decision trees. These improvements are even more prominent for lower b-tagging efficiencies.
Besides, the robustness of the result to some systematic effect such as pile-up
interaction, detector resolution, track reconstruction efficiency, and the training
sample size, is also investigated. About 20\% variation can be seen in the supposed 
scenarios and BDT with CTT always gives the best performance.
Also of note is that ANN with CTT shows a comparable performance to that of BDT without CTT.
Given these observations, boosted decision trees and cascade training should be seriously considered as an alternative to
artificial neural networks for tagging b-jets at collider experiments.
Note that the performance of both techniques may differ if other physics channels are considered, since the training
procedure may be affected by jet overlaps and gluon splitting into $b\bar{b}$ pairs.


\begin{acknowledgments}
We would like to thank Prof. I. Stancu and K. Mahn for reading the manuscript and for the
valuable comments.
This work has been supported by the US-DOE grant number DE-FG02-04ER46112 and by
Funda\c c\~ao para a Ci\^encia e Tecnologia under grant SFRH-BPD-20616-2004.
\end{acknowledgments}

\end{document}